\documentclass[11pt]{article}

\usepackage[margin=1in]{geometry}
\usepackage{graphicx}
\usepackage{booktabs}
\usepackage{amsmath}
\usepackage{siunitx}
\usepackage{xurl}
\usepackage{float}
\usepackage[colorlinks=true,citecolor=blue,linkcolor=blue,urlcolor=blue]{hyperref}
\usepackage[font=small, labelfont=bf, textfont=it, labelsep=period, justification=justified]{caption}

 \usepackage{cite}  % Compress multiple citations [1,2,3] -> [1-3]
 \usepackage{IEEEtrantools}

\title{%
	A Systems-Engineered ESP32 DAQ Architecture and FAIR Data Workflow for Small-Scale Wind Turbine Performance Measurement in Tropical Environments
}

\author{%
	A.~L.~Kulasekera\\
	Department of Mechanical Engineering, University of Moratuwa, Sri Lanka\\
	\texttt{asitha@uom.lk}\\
	ORCID: 0000-0002-6665-9209
}

\date{}

\begin{document}
	\bstctlcite{IEEEexample:BSTcontrol} % This activates the author limit
	\maketitle
	
	\begin{abstract}
		Small-scale wind turbine research in resource-constrained academic settings frequently produces unreliable or unpublishable datasets due to ad-hoc instrumentation, inadequate time synchronization, storage failures, and weak data governance. This paper presents a systematic data acquisition (DAQ) methodology and ESP32-based reference implementation design for field characterization of small wind turbines (100~W--5~kW), emphasizing tropical/coastal deployment constraints typical of Low- and Middle-Income Countries (LMIC). We integrate (i)~a student-adapted V-model with requirements traceability, (ii)~hardware selection strategies for high-humidity and salt-spray environments, (iii)~an embedded firmware architecture featuring interrupt-driven rotor speed measurement, state-machine fault handling, and NTP-based time synchronization, (iv)~a local-first hybrid storage design combining SD-card persistence with optional MQTT cloud telemetry, and (v)~a data-management workflow adapting CRISP-DM and FAIR principles with explicit quality dimensions and publication templates. A detailed helical vertical-axis wind turbine (VAWT) design scenario for coastal Sri Lanka illustrates the complete methodology, targeting $>90\%$ data completeness over six-month campaigns. The methodology is accompanied by open-source firmware, hardware templates, and data-publication workflow artifacts released via GitHub and Zenodo.
	\end{abstract}
	
	
	\section{Introduction}
	
	Small-scale wind energy systems (typically 100~W to 5~kW) represent essential distributed generation technologies for regions with limited grid infrastructure, particularly in tropical developing countries experiencing favorable monsoon wind regimes~\cite{bishop2008,chagas2020}. Rigorous performance characterization through field measurements, specifically the power coefficient (\(C_p\)) versus tip speed ratio (\(\lambda\)) curve, is fundamental for validating computational models, optimizing control strategies, and building empirical databases for future turbine design improvements. However, student researchers and early-career investigators face significant barriers: industrial standards (IEC 61400 series~\cite{iec61400-2-2013,iec61400-12-1-2017}, VDI 2206~\cite{vdi2206-2022}, ISO/IEC/IEEE 15288~\cite{isoiec15288-2023}) are comprehensive but overwhelming for time-limited projects; commercial SCADA systems are prohibitively expensive; and academic literature focuses on results rather than practical instrumentation methodology~\cite{heibati2023}. In tropical and coastal environments, high humidity, salt spray, and temperature extremes demand specialized equipment selections rarely addressed in temperate-climate guidance. Furthermore, mechanical and mechatronic engineering curricula often provide limited training in data-science practices, impeding proper data-quality controls and open data publication.
	
	\subsection{Problem Statement and Motivation}
	
	This work addresses the absence of an accessible yet rigorous, end-to-end methodology linking systems engineering (requirements, verification) to embedded implementation details (ISR timing, logging integrity, time synchronization) and to data governance (quality dimensions, FAIR publication) under resource constraints and tropical/coastal realities. Student projects commonly fail to reach publishable quality because measurement requirements are not made explicit, system architectures lack traceability, and post-processing proceeds without formal quality criteria or provenance documentation.
	
	\subsection{Contributions}
	
	We present the following methodological and engineering contributions:
	\begin{itemize}
		\item A student-adapted V-model systems engineering process~\cite{vdi2206-2022,gausemeier2002} with requirements traceability tailored to 6--12 month academic wind-turbine DAQ projects.
		\item Hardware design guidance for tropical/coastal deployments emphasizing enclosure protection (IP ratings), corrosion control (conformal coating, desiccant, stainless hardware), and maintainability within typical university budgets.
		\item An ESP32 firmware reference architecture with interrupt service routines (ISR) for rotor speed pulse counting, a fault-aware state machine, explicit derived-quantity computation (\(C_p\), \(\lambda\)), and NTP-based time synchronization~\cite{rfc5905}.
		\item A local-first hybrid storage design pattern: SD-card logging as the authoritative primary record with optional cloud/MQTT telemetry for monitoring and redundancy, robust to intermittent connectivity.
		\item A data-management workflow adapting CRISP-DM lifecycle stages~\cite{shimaoka2024,chapman2000} and FAIR principles~\cite{wilkinson2016} with defined data-quality dimensions, quality flags, and publication-ready metadata templates.
		\item A comprehensive helical VAWT design scenario for coastal Sri Lanka demonstrating the complete methodology application, including budget-constrained alternatives for LMIC contexts.
		\item Open-source release of firmware, calibration templates, and workflow artifacts via GitHub with archived release on Zenodo.
	\end{itemize}
	
	Companion resource: A more comprehensive educational guide (including step-by-step procedures, student-facing templates, and troubleshooting checklists) is provided in the companion book/report; this paper focuses on the research-style synthesis of the end-to-end DAQ-to-dataset methodology and reference architecture.
		
	\subsection{Paper Organization}
	
	Section~2 presents the systems engineering and measurement methodology. Section~3 describes the hardware and firmware reference architecture. Section~4 details the helical VAWT design scenario with standard and constrained budget alternatives. Section~5 presents the design targets and expected performance characteristics. Section~6 discusses transferability, limitations, and implementation considerations.
	
	\section{Methods and Design Framework}
	
	\subsection{Student-Adapted V-Model and Requirements Engineering}
	
\begin{figure}[H]
		\centering
		\includegraphics[width=1\linewidth]{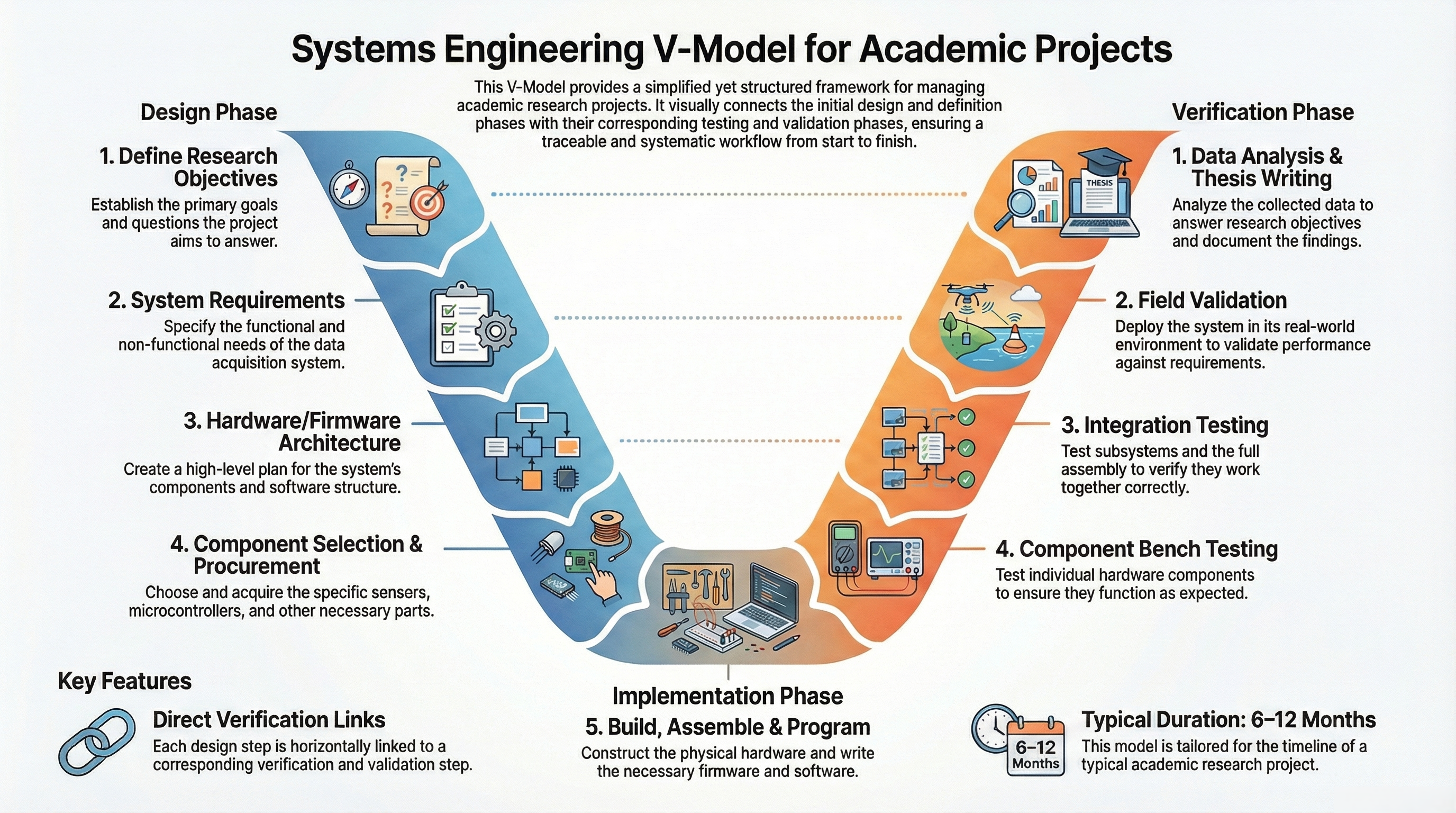}
		\caption{Simplified V-Model process adapted for student wind turbine DAQ projects. The left descending arm represents design decomposition from research objectives through system requirements to component specifications. The bottom vertex represents implementation (hardware assembly and firmware development). The right ascending arm represents verification and validation from component testing through field deployment to final data analysis. Horizontal dashed arrows indicate traceability between design decisions and verification activities. Typical timeline: 6--12 months for undergraduate/postgraduate projects.}
		\label{fig:vmodel}
	\end{figure}

	We employ a simplified V-model~\cite{vdi2206-2022,gausemeier2002} connecting requirements and architecture decisions (left arm) to verification and validation activities (right arm), tailored for student project constraints. Key artifacts remain lightweight, a concise requirements list, system architecture diagram, and verification checklist, while enforcing testability of each requirement (e.g., sustained sampling rate, timestamp integrity, minimum completeness target). Requirements cover measurement variables (wind speed, rotor speed, electrical power), operational metrics (uptime, data completeness), and data integrity (no duplicated timestamps, bounded clock drift, power-loss recoverability). A traceability matrix links each requirement to design elements (sensor selection, ISR timing model, buffering strategy) and verification methods (bench tests, 72-hour endurance runs, power-cycle tests, field shakedown). Figure~\ref{fig:vmodel} illustrates the adapted V-model framework for student DAQ projects.
	
	\subsection{System Architecture Decomposition}
	
\begin{figure}[b]
		\centering
		\includegraphics[width=1\linewidth]{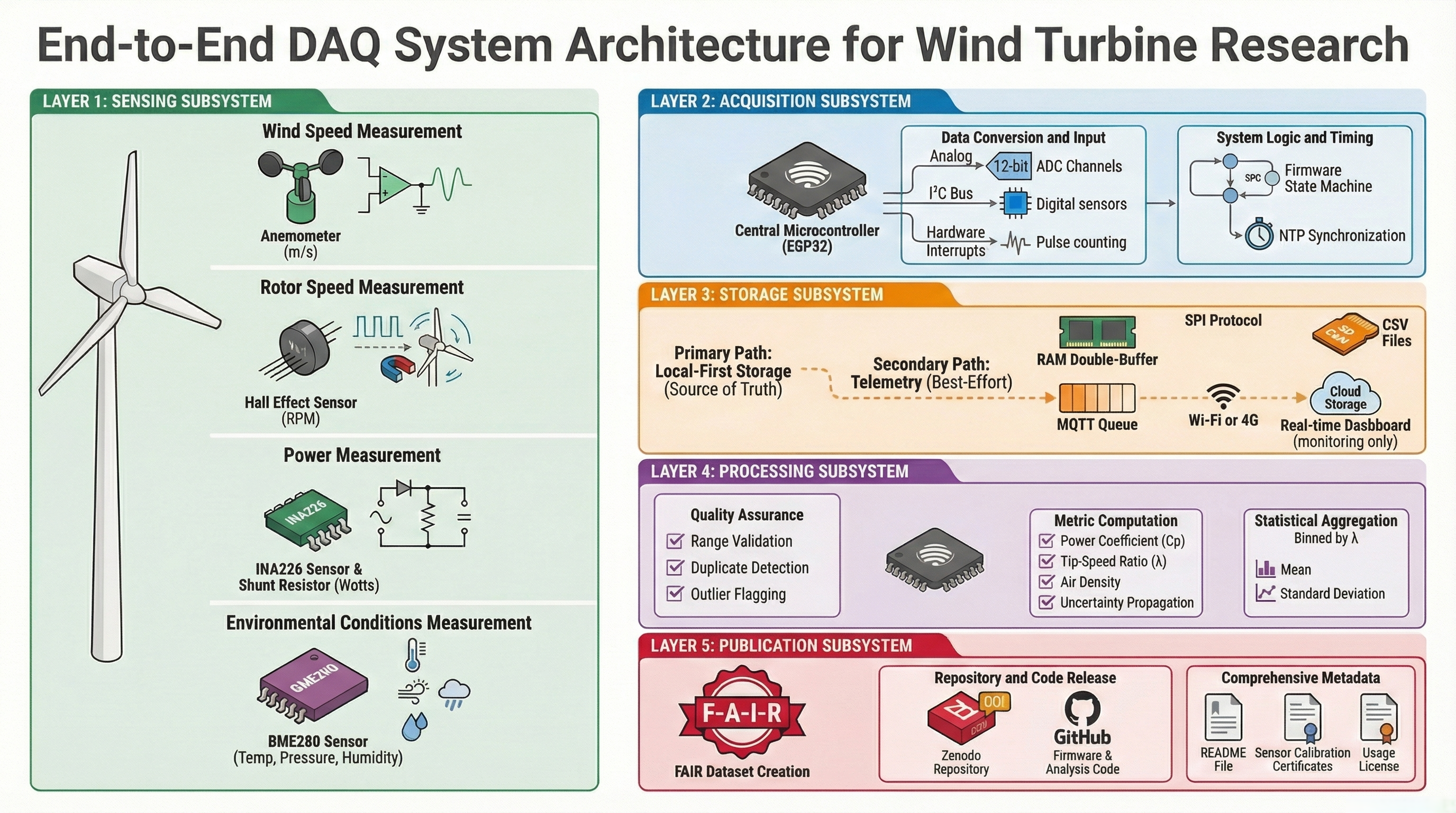}
		\caption{End-to-end DAQ system architecture: sensors (wind speed, rotor speed, power, environment) interface to ESP32 microcontroller with interrupt-driven acquisition and state-machine control; local-first SD persistence with optional cloud/MQTT telemetry; Python quality-check and performance-computation pipeline; FAIR-compliant dataset publication via Zenodo with companion GitHub resources.}
		\label{fig:architecture}
	\end{figure}

	The DAQ is decomposed into sensing, compute, storage, communication, and data-management subsystems with explicit interfaces and failure modes~\cite{zheng2014}. The architecture emphasizes local persistence and resilience to intermittent connectivity. Figure~\ref{fig:architecture} depicts the full pipeline from sensors through embedded processing to published dataset artifacts.
	
	\subsection{Tropical/Coastal Design Constraints}
	
	Tropical/coastal deployments introduce persistent humidity, salt spray, and corrosion affecting connectors, fasteners, PCB surfaces, and sensor bearings. The methodology treats environmental hardening as first-class requirements: IP-rated enclosures, cable glands with proper sealing, desiccant packs for moisture control, conformal coating (acrylic or silicone) for PCB protection, and corrosion-resistant mounting hardware (316-grade stainless steel). Maintenance assumptions, periodic inspection, cleaning, and desiccant replacement, are incorporated into the design methodology and inform expected data-quality outcomes.
	
	\subsection{Measurement Model and Uncertainty Propagation}
	
	The DAQ supports computation of wind-turbine performance metrics from measured wind speed \(v\), rotor speed \(\omega\), and power output \(P\). The power coefficient is defined as
	\[
	C_p = \frac{P_{\text{turbine}}}{\frac{1}{2} \rho A v^3},
	\]
	where \(\rho\) is air density (computed from temperature and pressure via environmental sensors), and \(A\) is swept area (for helical VAWTs, \(A = 2RH\)~\cite{hikkaduwa2012,sengupta2025}, where \(R\) is rotor radius and \(H\) is height). The tip speed ratio is
	\[
	\lambda = \frac{\omega R}{v}.
	\]
	Because \(C_p\) depends cubically on wind speed, calibration quality and uncertainty propagation are critical design considerations. The theoretical maximum \(C_p = 0.593\) (Betz limit~\cite{betz1920,manwell2009}) provides a physical constraint on achievable performance. Calibration procedures include two-point anemometer validation (wind tunnel or side-by-side field comparison), power sensor verification against reference multimeter, and environmental sensor ice-point/boiling-point checks. Combined uncertainty propagation using the Guide to the Expression of Uncertainty in Measurement (GUM) methodology~\cite{jcgm2008} is incorporated into the data-processing workflow.
	
	\section{Implementation}
	
	\subsection{Hardware Reference Architecture}
	
	The hardware stack centers on an ESP32 dual-core microcontroller with 12-bit ADC, Wi-Fi, and hardware interrupt capability. Sensor interfaces include:
	\begin{itemize}
		\item \textbf{Wind speed:} IEC-class anemometer (e.g., Inspeed Vortex) with \(\pm 0.3\)~m/s uncertainty at 10~m/s~\cite{iec61400-12-1-2017}, selected for accuracy-cost balance and tropical availability.
		\item \textbf{Rotor speed:} Hall-effect sensor (e.g., A3144) with neodymium magnets producing digital pulses suitable for hardware interrupts, ensuring sub-millisecond timing precision.
		\item \textbf{Power:} INA226 precision power monitor IC with external shunt resistor targeting \(\sim\)2\% accuracy, robust to field voltage transients.
		\item \textbf{Environment:} BME280 module (I\textsuperscript{2}C) measuring temperature, pressure, and humidity for air-density correction, housed in a multi-plate radiation shield.
	\end{itemize}
	Storage is implemented via industrial-grade SD card (SLC/MLC, 10,000+ write cycles) with SPI interface. Communication supports Wi-Fi with optional MQTT telemetry. Enclosures are IP67-rated polycarbonate with cable glands, conformal coating (MG Chemicals 422B silicone), and desiccant packs. Power supply includes 12~V AC adapter, sealed lead-acid battery backup, and LM2596 buck converters with reverse-polarity and surge protection. Table~\ref{tab:bom} summarizes key component selection rationale emphasizing tropical/coastal constraints.
	
	\begin{table}[t]
		\centering
		\caption{Condensed hardware decision matrix (standard budget scenario).}
		\label{tab:bom}
		\begin{tabular}{@{}lll@{}}
			\toprule
			Subsystem & Component & Rationale \\
			\midrule
			Wind speed & Inspeed Vortex & IEC-class accuracy, field-proven \\
			Rotor speed & Hall sensor + magnets & ISR-compatible digital pulses \\
			Power & INA226 + ext. shunt & 2\% accuracy target, transient-robust \\
			Compute/comms & ESP32-WROOM-32 & Dual-core, Wi-Fi, low cost \\
			Storage & 32~GB industrial SD & High endurance (10K cycles) \\
			Enclosure & IP67 polycarbonate & Humidity/salt-spray resilience \\
			Coating & Silicone conformal & Long-term coastal protection \\
			Hardware & 316 stainless steel & Corrosion resistance \\
			\bottomrule
		\end{tabular}
	\end{table}
	
	\subsection{Firmware Reference Architecture on ESP32}
	
	\subsubsection{ISR-Based Rotor Speed Measurement}
	
	Rotor speed is measured using a hardware interrupt service routine (ISR) attached to the Hall sensor output. The ISR increments a pulse counter with minimal latency (\(<1\)~\textmu{}s), decoupled from slower tasks (SD writes, network operations). A main-loop function computes RPM from pulse counts over a fixed time window, ensuring stable derived quantities without aliasing artifacts.
	
	\subsubsection{State Machine for Fault Handling}
	
\begin{figure}[b]
		\centering
		\includegraphics[width=1\linewidth]{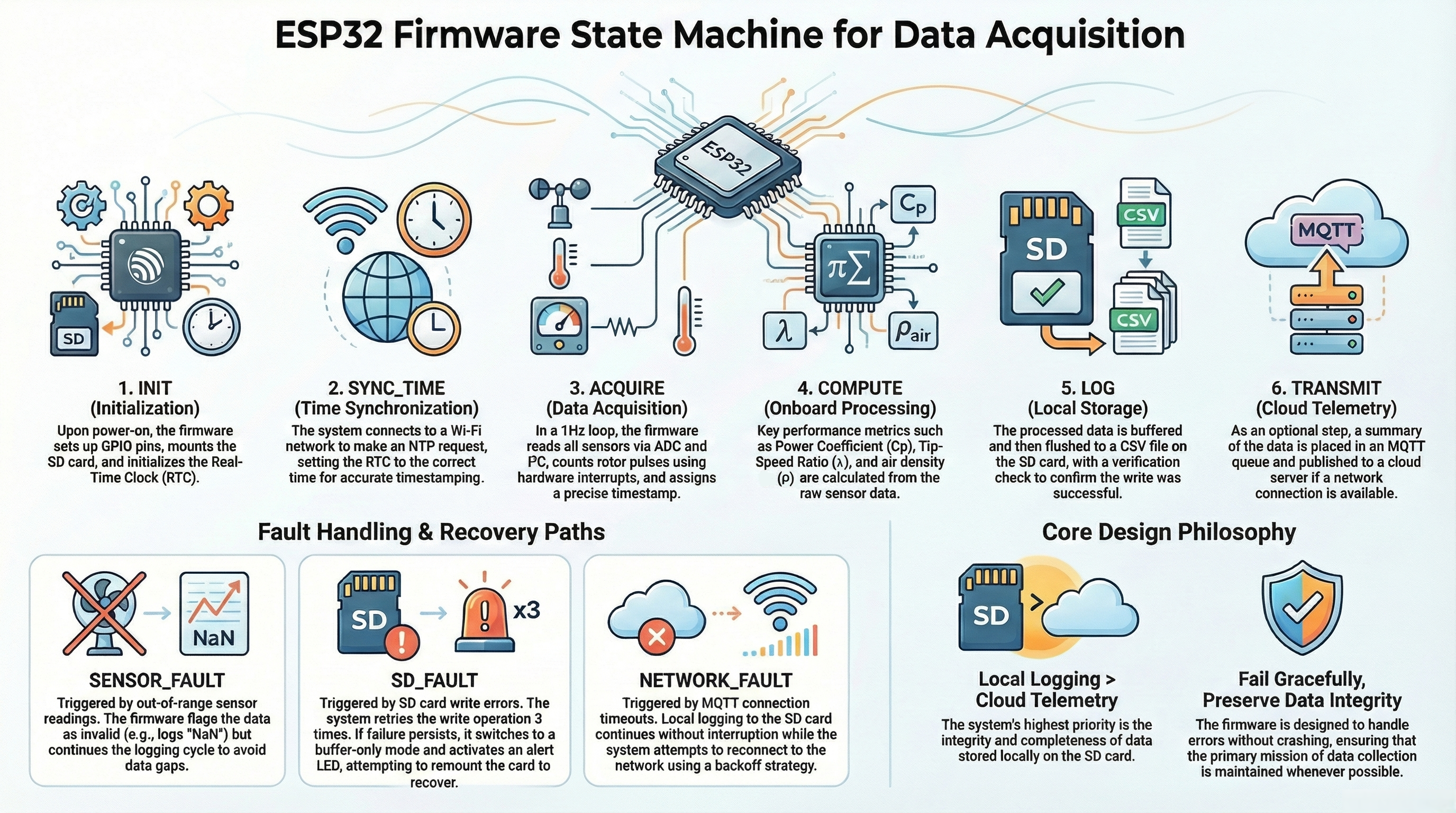}
		\caption{Firmware state machine with fault-aware transitions. Normal operation cycles through initialization, time synchronization (NTP), acquisition (1 Hz sampling), logging (SD write with buffering), and optional transmission (MQTT). Fault states handle sensor errors (flag and continue), SD failures (retry then buffer-only), and network outages (queue transmissions, local logging unaffected). Design prioritizes local data integrity over cloud connectivity.}
		\label{fig:statemachine}
	\end{figure}

	The firmware employs a state machine structuring initialization, acquisition, logging, connectivity, and error-handling behaviors. Fault states cover sensor invalid readings (e.g., out-of-range wind speed), SD-card write failures, and connectivity loss. Recovery actions prioritize local data integrity: SD logging continues during network outages; sensor faults trigger quality flags rather than system halts. This explicit fault model is designed to reduce silent failure modes during unattended operation. Figure~\ref{fig:statemachine} illustrates the fault-aware state machine architecture.
	
	\subsubsection{Sampling, Filtering, and Derived Computation}
	
\begin{figure}[b]
		\centering
		\includegraphics[width=1\linewidth]{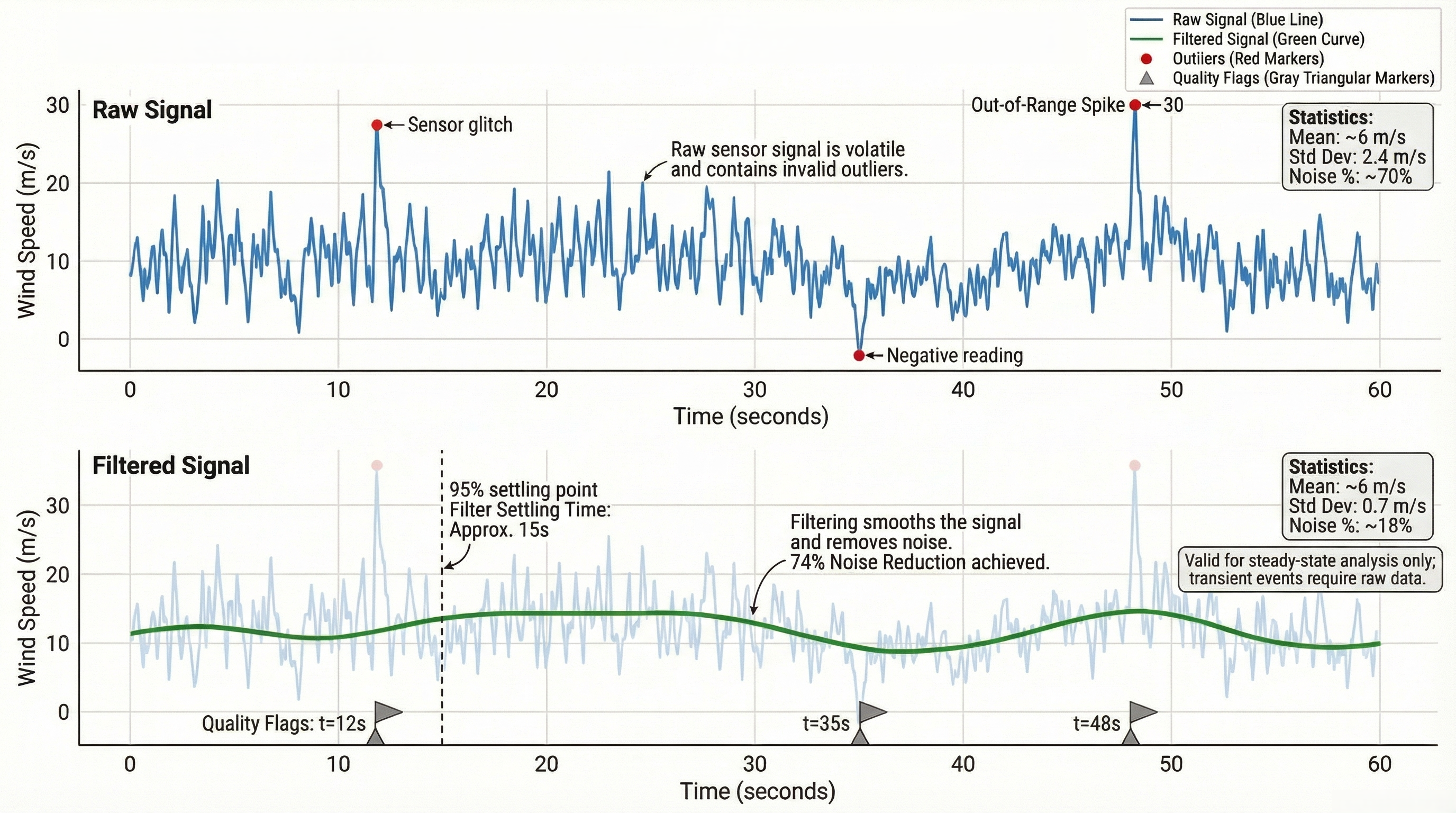}
		\caption{Example preprocessing effect: raw wind-speed samples (top) versus filtered/validated stream (bottom) used for derived metrics. Invalid/out-of-range samples (red markers) are flagged and excluded from steady-state binning. Exponential moving average filter ($\alpha = 0.2$) reduces measurement noise by approximately 70\% with 15-second settling time, suitable for 1 Hz wind turbine monitoring.}
		\label{fig:filtering}
	\end{figure}

	Acquisition proceeds at a fixed rate (typically 1~Hz sustained, configurable to 2~Hz for turbulent sites). Each sample includes raw sensor readings, timestamp, and computed derived quantities (\(C_p\), \(\lambda\), air density). Validation checks flag out-of-range values (e.g., wind speed \(> 25\)~m/s indicating sensor error, negative power indicating reverse current during startup) but preserve raw data for downstream analysis reproducibility. 	Figure~\ref{fig:filtering} demonstrates signal quality improvement through filtering and validation.
	
	\subsubsection{Time Synchronization (NTP and RTC)}
	
	Time synchronization uses Network Time Protocol (NTP)~\cite{rfc5905} over Wi-Fi to minimize clock drift. A hardware real-time clock (DS3231, \(\pm 2\)~ppm temperature-compensated) provides backup timekeeping during network outages and post-power-loss recovery. Timestamp integrity is enforced by detecting duplicated or out-of-sequence records, which are flagged as data-quality anomalies. Accurate timestamps are essential for binning performance curves and for cross-device correlation in multi-turbine studies.
	
	\subsubsection{Local-First SD Logging Strategy}
	
\begin{figure}[b]
		\centering
		\includegraphics[width=1\linewidth]{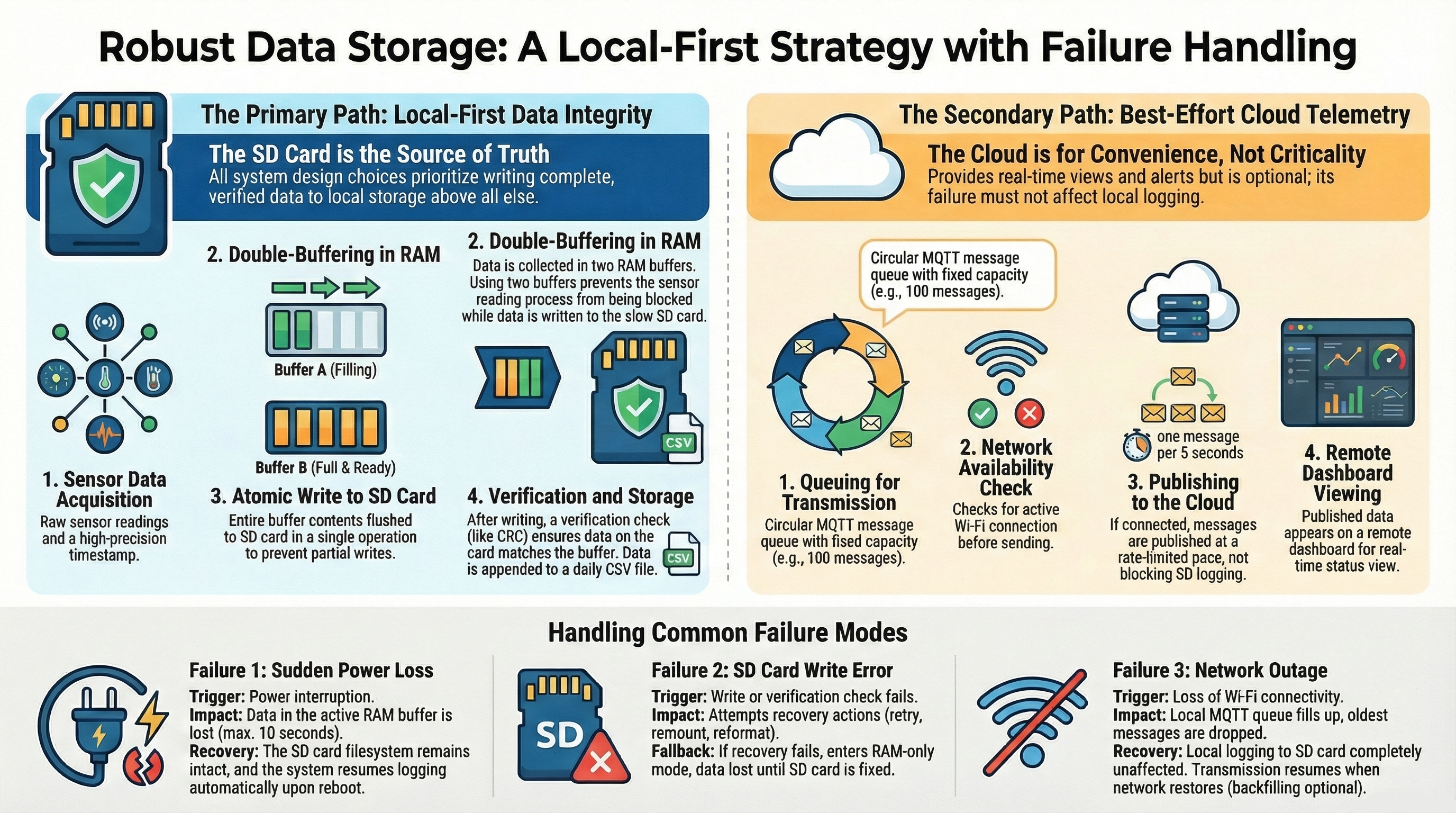}
		\caption{Local-first hybrid storage strategy: SD-card as primary source of truth with buffered writes; cloud upload as secondary for monitoring and redundancy when connectivity is available. Double-buffering and atomic flush operations protect against power-loss corruption. Network outages do not affect local logging; MQTT queue provides graceful degradation with exponential-backoff retry.}
		\label{fig:localfirst}
	\end{figure}

	SD-card logging is the authoritative primary record. Data is buffered in RAM and flushed atomically to minimize corruption risk during power interruptions. Double-buffering prevents acquisition blocking during SD writes. The firmware logs both raw sensor values and derived quantities in CSV format with headers, enabling direct analysis without binary parsing. SD integrity is verified by periodic read-back checks and file-system remounting after faults. Figure~\ref{fig:localfirst} illustrates the local-first storage architecture with failure-handling paths.
	
	\subsubsection{Hybrid Cloud Telemetry (MQTT)}
	
	Cloud communication via MQTT is optional, functioning as a secondary monitoring and redundancy channel. MQTT messages are queued and rate-limited to avoid destabilizing the sampling/logging path. The design handles intermittent Wi-Fi gracefully: failed transmissions are retried with exponential backoff, and local logging continues unaffected. Cloud telemetry enables real-time monitoring dashboards and alerts but is not required for data completeness or post-campaign analysis.
	
	\subsection{Data Pipeline: CRISP-DM Adaptation, Quality Dimensions, and FAIR Workflow}
	
\begin{figure}[b]
		\centering
		\includegraphics[width=1\linewidth]{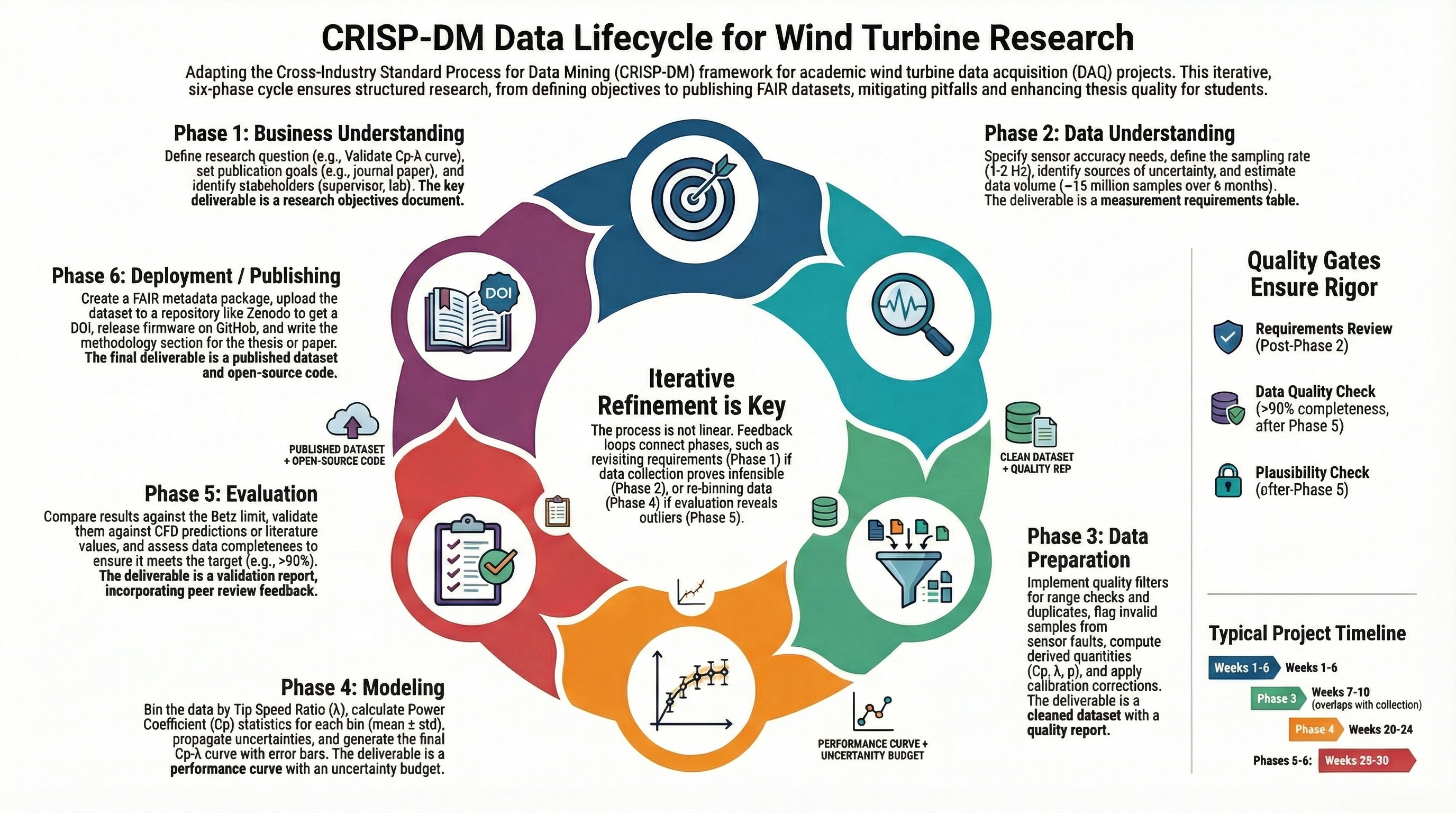}
		\caption{CRISP-DM workflow adapted for wind turbine DAQ projects. The methodology progresses from research objective definition through data acquisition design, quality-controlled preparation, performance curve modeling with uncertainty quantification, validation against physical constraints and prior work, to FAIR-compliant dataset publication. Iterative refinement between phases enables continuous quality improvement.}
		\label{fig:crispdm}
	\end{figure}

	The data workflow adapts Cross-Industry Standard Process for Data Mining  (CRISP-DM)~\cite{shimaoka2024,chapman2000} stages to a wind-turbine DAQ context.Figure~\ref{fig:crispdm} shows the adapted CRISP-DM workflow for wind turbine research. The key adapted stages include:
	\begin{itemize}
		\item \textbf{Business/Project Understanding:} Define research questions, performance metrics, and publication targets.
		\item \textbf{Data Understanding:} Specify sensor accuracy, sampling rate, expected operating ranges, and uncertainty contributors.
		\item \textbf{Data Preparation:} Implement quality filters (range checks, duplicate detection, sensor fault flags), compute derived quantities with uncertainty propagation~\cite{jcgm2008}.
		\item \textbf{Modeling/Analysis:} Bin \(C_p\)--\(\lambda\) data, compute statistics (mean, standard deviation per bin), generate performance curves with error bars.
		\item \textbf{Evaluation:} Validate against expected turbine behavior, compare to CFD predictions or prior literature.
		\item \textbf{Deployment/Publishing:} Package dataset with FAIR metadata~\cite{wilkinson2016}, README, calibration certificates, and open license.
	\end{itemize}
	
	Data quality is evaluated across multiple dimensions~\cite{dama2017}:
	\begin{itemize}
		\item \textbf{Completeness:} Percentage of expected samples successfully recorded (target \(>90\%\)).
		\item \textbf{Validity:} Conformance to physical ranges and sensor specifications.
		\item \textbf{Consistency:} Absence of duplicated or out-of-sequence timestamps.
		\item \textbf{Integrity:} Absence of SD corruption or data-loss events.
		\item \textbf{Timeliness:} Bounded clock drift and synchronization accuracy.
	\end{itemize}

	FAIR (Findable, Accessible, Interoperable, Reusable) publication~\cite{wilkinson2016} is implemented as a concrete package:
	\begin{itemize}
		\item \textbf{Findable:} Zenodo DOI, descriptive metadata (title, author, keywords, site coordinates, turbine specs, deployment dates).
		\item \textbf{Accessible:} Open download without authentication; dataset README with usage instructions.
		\item \textbf{Interoperable:} CSV format with SI units, standard variable names, and metadata schema.
		\item \textbf{Reusable:} Open license (CC-BY-4.0), calibration certificates, uncertainty documentation, firmware version tag.
	\end{itemize}
	
	\section{Design Scenario: Helical VAWT for Coastal Sri Lanka}
	
	\subsection{Scenario Context}
	
	A comprehensive helical vertical-axis wind turbine (VAWT) DAQ design scenario for coastal Sri Lankan deployment demonstrates the complete methodology. Two budget scenarios are developed: (i)~standard budget (approximately 1~million~LKR or 2,500~USD for DAQ hardware) targeting journal-grade data quality, and (ii)~constrained budget (300,000~LKR or 750~USD) using consumer-grade sensors with rigorous calibration strategies, suitable for LMIC undergraduate projects targeting conference-grade datasets.
	
	\subsection{Standard Budget Configuration}
	
	The standard-budget configuration specifies the IEC-class anemometer with two-point wind-tunnel calibration protocol (0~m/s and 10~m/s reference points)~\cite{iec61400-12-1-2017}, INA226 power measurement with validation procedure against a Fluke 87V multimeter at multiple test points, BME280 environmental sensor calibration at ice-point and boiling-point, and Hall sensor rotor speed measurement verification via optical tachometer cross-check. The design includes complete calibration workflow documentation, correction factor computation templates, and dataset package specifications.
	
	\subsection{Constrained Budget Alternative}
	
	The constrained-budget alternative employs consumer-grade anemometer (Davis Instruments, $\pm$ 1.0~m/s stated accuracy) with monthly side-by-side field calibration against a borrowed reference instrument, single INA226 module without external precision shunt (accepting 3--4\% accuracy), and DIY environmental protection measures (3D-printed radiation shield, IP65 enclosure with silicone sealing, galvanized steel hardware with annual replacement expectation). This scenario demonstrates that systematic design and rigorous calibration workflows can enable meaningful research outcomes under severe budget constraints. Table~\ref{tab:scenarios} compares the two budget scenarios.
	
	\begin{table}[t]
		\centering
		\caption{Budget scenario comparison for helical VAWT DAQ system (coastal Sri Lanka).}
		\label{tab:scenarios}
		\small
		\begin{tabular}{@{}p{0.22\linewidth}p{0.32\linewidth}p{0.32\linewidth}@{}}
			\toprule
			\textbf{Parameter} & \textbf{Standard Budget (\$2,500)} & \textbf{Constrained (\$750)} \\
			\midrule
			Wind speed sensor & IEC-class anemometer, $\pm 0.3$ m/s & Consumer-grade, $\pm 1.0$ m/s, monthly calibration \\
			Power measurement & INA226 + external shunt, 2\% & INA226 only, 3--4\% \\
			Enclosure & IP67 polycarbonate & IP65 modified junction box \\
			Hardware & 316 stainless steel & Galvanized steel, annual replacement \\
			Calibration & Wind tunnel 2-point & Side-by-side field comparison \\
			Expected completeness & 92--95\% & 85--90\% \\
			Data quality grade & Journal-grade & Conference-grade \\
			\bottomrule
		\end{tabular}
	\end{table}

	\subsection{Operations and Maintenance Model}
	
	The design methodology specifies an operations model assuming unattended operation with monthly maintenance visits: visual inspection, SD-card backup download, anemometer cleaning (salt accumulation), and quarterly tasks including calibration drift checks and desiccant replacement. The firmware's fault handling and local-first storage design are architected to minimize irreversible data loss during outages. Campaign duration target is six months, with expected data completeness targets differentiated by budget scenario.
	
	\section{Design Targets and Expected Performance}
	
	\subsection{Expected Data Quality and Completeness}
	
	The standard-budget scenario targets the following performance characteristics based on component specifications and design analysis:
	\begin{itemize}
		\item Data completeness: 90--95\% over six-month campaigns (accounting for monthly maintenance downtime, occasional power outages with battery backup, and sensor cleaning intervals).
		\item System uptime: \(>95\%\) with battery backup surviving typical 4--6 hour power outages.
		\item Sustained sampling rate: 1~Hz continuous, expandable to 2~Hz for high-turbulence sites.
		\item SD corruption mitigation: double-buffered writes with atomic flush and integrity verification designed to prevent data loss.
		\item Coastal survival: IP67 enclosure, conformal coating, and corrosion-resistant hardware designed for \(>6\) month deployment without component failures.
	\end{itemize}
	
	Table~\ref{tab:targets} summarizes design targets for the standard-budget scenario.
	
	\begin{table}[t]
		\centering
		\caption{Design targets and expected performance (standard budget scenario).}
		\label{tab:targets}
		\begin{tabular}{@{}ll@{}}
			\toprule
			Parameter & Target Value \\
			\midrule
			Campaign duration & 6 months \\
			Expected data completeness & 90--95\% \\
			Sampling rate (sustained) & 1~Hz (up to 2~Hz) \\
			Target system uptime & >95\% \\
			Wind speed uncertainty & $\pm 0.3$~m/s at 10~m/s \\
			Power measurement uncertainty & $\sim$2\% \\
			Combined $C_p$ uncertainty & 5--6\% (propagated) \\
			Maintenance interval & Monthly (visual + download) \\
			Calibration check interval & Quarterly \\
			\bottomrule
		\end{tabular}
	\end{table}
	
	\subsection{Quality Assurance and Verification Strategy}
	
	The methodology specifies explicit verification activities:
	\begin{itemize}
		\item \textbf{Bench testing:} 72-hour continuous acquisition run with simulated sensor inputs, verifying data integrity, power-cycle recovery (50 iterations), and thermal stability in enclosed conditions.
		\item \textbf{Sensor calibration:} Two-point anemometer validation~\cite{iec61400-12-1-2017}, power sensor multi-point verification, environmental sensor reference-point checks with documented calibration certificates.
		\item \textbf{Field shakedown:} One-week commissioning period verifying sensor installation, checking initial data completeness (\(>95\%\) target), and adjusting sampling parameters if needed.
		\item \textbf{Quality filters:} Post-processing scripts applying range checks (wind speed \(<0\) or \(>25\)~m/s, rotor RPM \(<0\) or \(>500\)), duplicate timestamp detection, and sensor fault flag generation.
	\end{itemize}
	
	Expected data retention after quality filtering: 92--95\% of raw samples pass validity checks under normal operating conditions.
	
	\subsection{Reproducibility Resources}
	
	The methodology is accompanied by open-source resources:
	\begin{itemize}
		\item \textbf{GitHub repository:} Tested ESP32 firmware (Arduino/PlatformIO), calibration spreadsheet templates, data-quality Python scripts, hardware wiring diagrams (Fritzing/KiCad), and troubleshooting flowcharts (\url{https://github.com/asithakal/wind-turbine-daq-guide}).
		\item \textbf{Zenodo archival release (v1.1):} Tagged firmware and template release with persistent DOI (\url{https://doi.org/10.5281/zenodo.18093662}), enabling version-controlled reproducibility.
		\item \textbf{Documentation artifacts:} Requirements templates, BOM templates with sourcing guidance, dataset README templates conforming to FAIR metadata standards~\cite{wilkinson2016}, and complete calibration workflow procedures.
	\end{itemize}
	
	\section{Discussion}
	
	\subsection{Transferability and Adaptation}
	
	The methodology is designed for transferability across small-wind configurations (HAWT/VAWT) with explicit guidance on swept-area definitions~\cite{hikkaduwa2012,sengupta2025}, calibration requirements, and uncertainty propagation appropriate to different turbine architectures. The V-model approach with requirements traceability~\cite{vdi2206-2022} remains applicable regardless of specific sensor choices, while the local-first storage pattern and FAIR publication workflow~\cite{wilkinson2016} are sensor-agnostic. The constrained-budget scenario demonstrates that systematic engineering practices combined with rigorous calibration can expand accessibility for LMIC institutions where typical commercial DAQ solutions are cost-prohibitive.
	
	\subsection{Implementation Considerations}
	
	Successful implementation requires several preconditions:
	\begin{itemize}
		\item \textbf{Institutional support:} Laboratory facilities for bench testing, access to calibration equipment (wind tunnel or reference instruments), and technical staff consultation.
		\item \textbf{Site preparation:} Turbine installation with appropriate tower/mounting, electrical power availability (with backup recommended), and network connectivity (Wi-Fi or cellular) if cloud telemetry is desired.
		\item \textbf{Maintenance commitment:} Realistic assessment of maintenance cadence achievable by the research team; the design targets assume monthly visits and quarterly calibration checks.
		\item \textbf{Timeline alignment:} The methodology is structured for 6--12 month academic project cycles; shorter timelines may require scope reduction.
	\end{itemize}
	
	Field completeness and quality outcomes will vary with site-specific power stability, environmental extremes (e.g., monsoon intensity, lightning exposure), and the maintenance discipline maintained by the team.
	
	\subsection{Limitations and Out-of-Scope Items}
	
	This work does not address industrial SCADA integration, utility-scale wind farm monitoring, or commercial certification processes. It excludes grid interconnection and power quality analysis (IEC 61400-21 scope~\cite{iec61400-21-2008}), advanced CFD development, structural dynamics and FEA, and detailed aerodynamic blade optimization. The focus remains on instrumentation methodology, data integrity design, and publishable dataset preparation workflows for academic research contexts.
	
	\subsection{Future Validation Work}
	
	While the design scenario presented here is comprehensive and grounded in component specifications, field deployment validation remains necessary to confirm expected performance characteristics under real tropical/coastal conditions. Future work should report measured completeness, uptime, and data-quality outcomes from actual deployments, quantify failure modes and maintenance burden, and refine calibration intervals based on observed sensor drift rates. Additionally, expanding the methodology to multi-turbine installations and grid-connected scenarios would broaden applicability.
	
	\section{Conclusion}
	
	This paper presents a systematic, reproducible DAQ-to-dataset methodology for small wind turbine field research integrating a tailored V-model with requirements traceability~\cite{vdi2206-2022}, a robust ESP32 firmware reference architecture, a local-first hybrid logging design pattern, and a CRISP-DM~\cite{shimaoka2024} + FAIR~\cite{wilkinson2016}-aligned data governance workflow. The comprehensive helical VAWT design scenario for coastal Sri Lanka demonstrates the methodology's application with detailed component selection rationale, calibration procedures, and budget-constrained alternatives suitable for LMIC contexts. The companion open-source resources, firmware, templates, and workflow artifacts, are positioned to reduce implementation barriers and raise baseline reproducibility for small-wind studies in tropical/coastal environments and resource-constrained academic settings. Field validation of the design targets remains an important direction for future work.
	
%	\section*{Acknowledgements}
%	The author thanks the University of Moratuwa Department of Mechanical Engineering for support in developing this methodology and educational resource.
	
	\section*{Funding and Conflicts of Interest}
	This work received no external funding. The author declares no conflicts of interest.
	
%	\section*{Conflicts of Interest}
%	The author declares no conflicts of interest.
	
	\section*{Code Availability}
	Firmware, templates, and methodology documentation: \url{https://github.com/asithakal/wind-turbine-daq-guide}. 
	Archived release (v1.1): \url{https://doi.org/10.5281/zenodo.18093662}.
	
	% Bibliography - using plain style for numerical citations
	\bibliographystyle{IEEEtran}
	\bibliography{references}

%	\printbibliography
	
\end{document}